\documentclass[lettersize,journal]{IEEEtran}

\usepackage{cite}
\usepackage{amsmath,amssymb,amsfonts}
\usepackage{algorithmic}
\usepackage{graphicx}
\usepackage{textcomp}
\usepackage{xcolor}
\usepackage{gensymb} 
\usepackage{array} 
\usepackage{stfloats}
\usepackage{url}
\usepackage{cuted}      
\usepackage{capt-of}    
\usepackage{verbatim}
\usepackage{subfigure}
\usepackage{rotating}   
\usepackage{float}
\usepackage{booktabs}
\usepackage{multirow}
\usepackage[utf8]{inputenc}

\usepackage{hyperref}
\hypersetup{hidelinks=true}

\def\BibTeX{{\rm B\kern-.05em{\sc i\kern-.025em b}\kern-.08em
    T\kern-.1667em\lower.7ex\hbox{E}\kern-.125emX}}

\title{A Synchronized Tri-Modality Wireless System for Multi-Point Non-Contact Cardiac Monitoring}

\author{Wei Ren, \IEEEmembership{Member, IEEE}, Kaili Zhao, Huansheng Yi, Miaoyang Hu, \\ 
Jianqi Wang, Canhua Xu, and Fugui Qi, \IEEEmembership{Member, IEEE}
\thanks{Wei Ren and Kaili Zhao contributed equally to this work. This work was supported in part by the National Natural Science Foundation of China under Grant 62201578; and in part by Joint Founding Projectof Innovation Research Institute, Xijing Hospital (LHJJ24YG14).
\textit{(Corresponding author: Fugui Qi.)}}
\thanks{Wei Ren is with Hong Kong Generative AI Research and Development Center, Hong Kong University of Science and Technology, Hong Kong SAR (email: weiren\_thu@163.com).}
\thanks{Kaili Zhao is with the Nursing Department, The Third Affiliated Hospital of Southern Medical University, Guangzhou 510630, China (email: zhaokaili1991@163.com).}
\thanks{Huansheng Yi, Jianqi Wang, and Canhua Xu are with the School of Biomedical Engineering, Fourth Military Medical University, Xi'An 710032, China (email: Yihsbme@outlook.com, wangjq@fmmu.edu.cn, canhuaxu@fmmu.edu.cn).}
\thanks{Miaoyang Hu is with the Department of Cardiology, Xijing Hospital, Fourth Military Medical University, Xi'an, 710032, China (email: humiaoyang@qq.com).}
\thanks{Fugui Qi is with the School of Biomedical Engineering, Fourth Military Medical University, and Military Medical Innovation Center, Fourth Military Medical University, Xi'An 710032, China (email: Qifgbme@outlook.com).}
}

\begin{document}

\maketitle

\bstctlcite{IEEEexample:BSTcontrol}

\begin{abstract}
Cardiac physiological monitoring is fundamental to evaluating cardiovascular health and diagnosing related diseases.
Non-contact radar sensing can capture chest-wall vibrations with spatial specificity, enabling localized physiological monitoring and cardiac assessment.
Progress, however, has been limited by the absence of a reliable, research-grade hardware platform for rigorous validation of radar-based methods.
Prior systems typically pair radar, namely Doppler cardiography (DCG), with only electrocardiography (ECG) or other coarse references, providing no direct mechanical ground truth and thus constraining fine-grained evaluation of radar outputs.
We present a tri-modality wireless system that integrates ECG, seismocardiography (SCG), and DCG under a unified acquisition and control framework for time-aligned recording and analysis.
The tri-modality system combines miniaturized wearable ECG/SCG modules with a high-resolution mmWave radar module, supporting spatially resolved sensing and subsequent feature extraction.
To our knowledge, this is the first integrated ECG--SCG--DCG hardware system for comprehensive cardiac monitoring;
it enables multi-point thoracic acquisition, achieves precise cross-modal synchronization, and maintains specifications suitable for wearable operation.
Experiments on a linear-motion platform show close agreement between DCG- and SCG-derived displacements, confirming accurate displacement measurement.
Human-subject studies further demonstrate cross-modal consistency and robust extraction of feature points, establishing ECG as the electrical anchor and SCG as the mechanical ground truth for DCG.
Overall, the system provides a rigorous, spatially resolved validation platform that advances non-contact cardiac sensing and supports next-generation radar-based monitoring.
\end{abstract}


\section{Introduction}
\label{sec:introduction}

Cardiac physiological monitoring is fundamental to assessing cardiovascular health and diagnosing related disease \cite{WHO2025CVD}. 
Cardiac function can be characterized through two coupled activities: the heart's electrical activity and mechanical vibrations. 
Electrical activity is routinely measured by electrocardiography (ECG), which provides high temporal resolution markers of depolarization--repolarization (P--QRS--T), robust R-peak timing for cycle delineation, and standardized acquisition/interpretation protocols that establish ECG as the clinical reference for rhythm analysis and heart-rate estimation \cite{pan2007real,kligfield2007recommendations}. 
However, ECG requires specialized instrumentation and contact-based measurement with skin electrodes, which is inconvenient in many settings.
Mechanical activity is captured by seismocardiography (SCG) using chest-mounted accelerometers; 
SCG reflects thoracic micro-vibrations driven by cardiac mechanics and enables extraction of key feature points---such as aortic-valve opening (AO), aortic-valve closure (AC), mitral-valve closure (MC), and mitral-valve opening (MO)~\cite{lin2023_scg_gmm_jbhi,zia2020_scg_quality_jbhi,zanetti2013seismocardiography,mora2020_sca_annotation_tim,cocconcelli2020_scg_annotation_tim}. 
In addition, because these surface vibrations originate from distinct cardiac chambers and pulmonary regions, they exhibit intrinsic spatial specificity even within a single cycle, enabling localized and more informative assessment of underlying physiology~\cite{crow1994relationship}.
Recent advances extend the spatially localized mechanical sensing to non-contact modalities, most notably millimeter-wave (mmWave) radar--based Doppler cardiography (DCG), which infers chest-wall micro displacements without electrodes or skin contact, thereby improving comfort and hygiene while enabling spatially specific measurement of cardiac mechanics as a complement to ECG 
\cite{eder2023_fmcw_sparsity_jbhi,Ye2019TBME_RadarSG,Hu2014TBME_DopplerRadar,zhang2019bioradar,dong2024review,ren2024noncontact,adib2015smart,ren2021vital,gu2013hybrid,dai2021enhancement,li2021through,gu2010_instrument_tim,bakhtiari2012_mmwave_tim,zhao2017_digitalif_tim,xia2021_cardiac_tim,zhang2023_overview_tim,liMotionRobustContactlessHeartbeat2023a}.

Despite this promise, progress has been constrained by the lack of a reliable, research-grade platform for rigorous validation of radar-based methods. 
Most current radar-based physiological monitoring and further diagnosis follow a two-stage paradigm: chest-wall displacement or cardiac-motion waveforms are first estimated from radar echoes, and physiological status or cardiac events are then inferred from the estimated motion signals \cite{yuan2025atrial,dong2024robust,chen2022contactless,paterniani2023radar,dong2024review,xia2021_cardiac_tim,zhang2023_overview_tim}.  
Recent TIM studies have substantially improved radar-based cardiac sensing from different perspectives, including high-accuracy cardiac activity extraction, 4-D imaging-radar heartbeat sensing, noncontact ECG reconstruction, and vital-sign estimation under random body movement~\cite{xia2021_cardiac_tim,liMotionRobustContactlessHeartbeat2023a,li2024_radarnet_tim,qiao2025_missing_tim}. 
The limited performance of existing methods, however, still largely depends on the accuracy and reliability of displacement estimation in the first step.
Most existing systems pair DCG with only ECG or other coarse sensors, such as breathing belts and handheld pulse oximeters~\cite{ren2021vital,sakamotoNoncontactMeasurementHuman2019,liMotionRobustContactlessHeartbeat2023a,li2024_radarnet_tim}.
While ECG provides an accurate electrical anchor and precise temporal markers (e.g., R-peaks), it is not a mechanical ground truth for radar-based displacement. 
As a result, fine-grained evaluation of DCG-derived displacement and cardiac feature points remains limited. 
A validation platform for this application should (i) include a clinical electrical reference (ECG) for timing, cycle delineation, and physiological-status reference, 
(ii) include a direct mechanical reference such as SCG to serve as ground truth for radar-derived displacement, and 
(iii) support multi-point, highly synchronized, and convenient data acquisition across the thorax.

To address these needs, we develop a tri-modality wireless system that integrates ECG (electrical), SCG (mechanical), and DCG (radar) within a unified acquisition and control framework to ensure time-aligned recording and analysis. 
The platform combines miniaturized wearable ECG/SCG modules with a high-resolution mmWave radar module, enabling multi-point thoracic measurements for spatial analysis and subsequent feature extraction. 
It achieves precise cross-modal synchronization while maintaining specifications suitable for wearable operation.

We demonstrate the performance of this platform in benchtop and human-subject studies. 
On a programmable linear-motion bench, DCG- and SCG-derived displacements show close agreement (e.g., low root-mean-square error and high Pearson correlation), validating accurate displacement tracking. 
In human subjects, the system enables robust extraction of cardiac feature events---including AO and MO---with cross-modal consistency; 
within this framework, ECG serves as the electrical anchor and SCG provides the mechanical ground truth for DCG. 
Overall, these results establish a rigorous, spatially resolved validation platform that advances non-contact cardiac sensing and supports next-generation radar-based monitoring.

The remaining sections of this paper are organized as follows. 
Section~\ref{sec:hardware system} details the design of the tri-modality hardware system. 
Section~\ref{sec:data processing methods} describes the methodology for data processing and feature point extraction methods. 
Section~\ref{sec:experiment and results} reports benchtop and human-subject experiments that validate system performance.
Section~\ref{sec:conclusion} concludes the paper. 
\section{System Architecture and Hardware Design}
\label{sec:hardware system}

\subsection{System Overview}

\begin{figure*}[htbp]
\centerline{\includegraphics[scale=0.95]{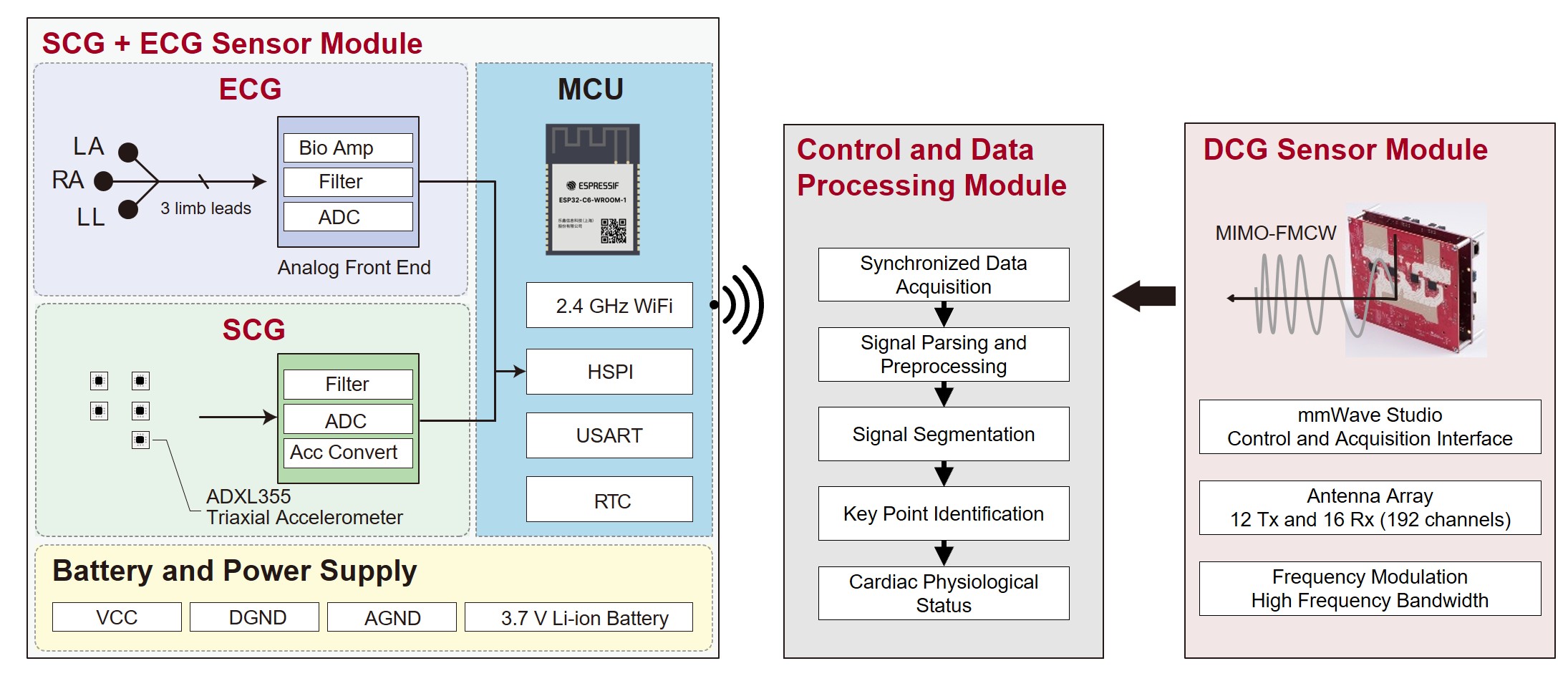}}
\caption{Tri-modality platform architecture. The system comprises: (a) ECG as the electrical anchor; (b) SCG as the direct mechanical reference via multi-point chest accelerometry; (c) mmWave DCG for non-contact chest micro-displacement; and (d) a unified acquisition/control/processing module that provides synchronized, cross-modal operation for spatially resolved cardiac monitoring.}
\label{fig:MultiVital hardware system}
\end{figure*}

\begin{figure}[htbp!]
\centerline{\includegraphics[scale=0.16]{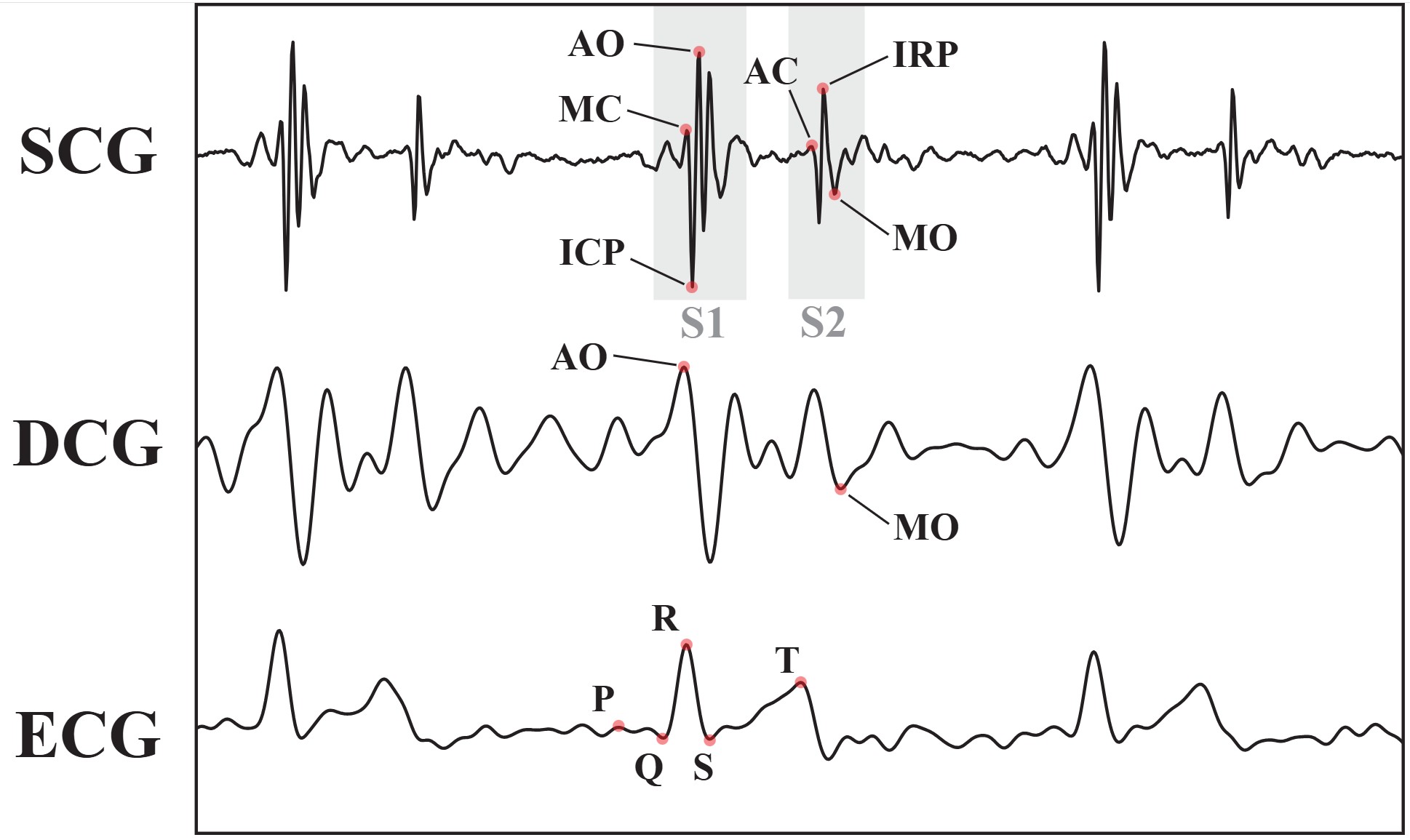}}
\caption{Synchronized ECG–SCG–DCG waveforms and key feature points. ECG provides precise R-peak timing for cycle delineation; SCG displays the systolic (S1) and diastolic (S2) complexes with key feature points MC/ICP/AO and AC/IRP/MO, respectively; DCG provides a radar-derived mechanical surrogate recorded concurrently. }
\label{fig:SCG DCG and ECG signals}
\end{figure}

The primary function of this system is to synchronously record both the subtle mechanical vibrations on the thoracic surface caused by cardiac activity, and the concurrently generated physiological electrical signals. 
The overall system design is illustrated in Fig.~\ref{fig:MultiVital hardware system}.
The system comprises the following components:
1) SCG and ECG sensor module. These modules are for directly acquiring thoracic micro-vibration signals (SCG) and electrical signals (ECG). 
A total of five SCG sensors are deployed to enable multi-point signal acquisition across the chest.
2) mmWave radar (DCG) module. This module is for non-contact acquisition of dynamic chest wall displacement signals.
3) Data acquisition and control module. This is for managing all three sensor modalities. The software supports synchronized initiation and termination of signal acquisition, data download, and subsequent signal processing and analysis.

\subsubsection{SCG and ECG Modules}



The primary function of these modules are to enable wireless acquisition of ECG and SCG signals from a human subject using customized miniaturized wearable sensor array.
The sensor array consists of five SCG sensors and one ECG sensor, attached at designated positions on the thoracic surface. 
The collected signals are transmitted to a host computer for subsequent processing.

As illustrated in Fig.~\ref{fig:SCG DCG and ECG signals}, a representative lead II ECG signal from a healthy human subject is shown. 
A complete cardiac cycle in the ECG consists of the P wave, QRS complex, and T wave. The R wave typically exhibits the highest amplitude. 
The R wave is therefore used as a temporal reference point for segmenting heartbeat cycles during data processing.
Fig.~\ref{fig:SCG DCG and ECG signals} also presents a SCG signal synchronously recorded with the ECG. 
The SCG waveform comprises two major components: the systolic complex (S1) and the diastolic complex (S2).
The S1 complex includes three prominent feature points: MC, isovolumetric contraction point (ICP), and AO; while the S2 complex includes AC, isovolumetric relaxation point (IRP), and MO.
These six feature points are clearly identifiable in SCG signals and serve as key reference markders for subsequent multimodal signal analysis. 

\subsubsection{DCG Module}
This module integrates an mmWave radar system to noninvasively acquire DCG signals.
This radar system consists of two key components: MMWCAS-RF-EVM for radar waveform generation and echo reception, and MMWCAS-DSP-EVM for data acquisition and transfer \cite{TI_MMWCAS_RF_EVM,MMWCAS-DSP-EVM}.
The system operates by emitting electromagnetic waves into the surrounding environment, and collects the signals reflected back by the human body and surrounding objects. 
The reflected signals are subsequently processed with advanced signal processing techniques to determine the individual's physiological condition.
This radar system employs a multiple-input multiple-output (MIMO) antenna array to enhance angular resolution and achieve fine-grained detection.
Fig.~\ref{fig:SCG DCG and ECG signals} also displays a representative DCG signal that was synchronously acquired alongside the ECG.

\subsubsection{Data Acquisition and Control Module}
This module provides centralized control of SCG and ECG sensor modules, and DCG module. 
Through wireless communication, it issues commands to initiate and terminate data acquisition simultaneously across all sensor modules, ensuring time-synchronized recording.
This module also facilitates wireless data retrieval from each sensor module after data acquisition is complete. 
The collected data from all modules are downloaded and stored locally on the host computer. 
Subsequently, the stored data undergo a series of processing steps, including data parsing and signal preprocessing, signal segmentation, SCG key point extraction, and estimation of cardiac physiological parameters.

The primary specifications of the overall system and the three sensor modules are summarized in Table.~\ref{tab:system parameter configurations}.
The detailed configurations and design considerations are described in the following sections.


\begin{table}[]
\caption{System Configurations and Per-modality Specifications.}
\centering
\begin{tabular}{llr}
\cline{1-3}
\multicolumn{2}{l}{\textbf{Parameter}}  & \textbf{Value} \\ 
\cline{1-3}

\multicolumn{2}{l}{Signal Modalities}                   & ECG, SCG, DCG     \\
\multicolumn{2}{l}{Totoal System Power}                   & $0.74~\text{W}$     \\
\multicolumn{2}{l}{Battery Capacity}                   & $700~\text{mAh}$     \\
\cline{1-3}
\multirow{4}{*}{ECG} 	      &{Sensor Demensions}                   & $3~\text{cm}\times 3~\text{cm}\times 1.5~\text{cm}$     \\
                            & Sampling Rate 					& $150~\text{Hz}$     \\
                   		& Resolution (Bit Depth) 	& $20~\text{bit}$     \\
                   		& Bandwidth 							& $0.01-30~\text{Hz}$ \\
\cline{1-3}
\multirow{4}{*}{SCG} 	      &{Sensor Demensions}                   & $3~\text{cm}\times 3~\text{cm}\times 1.5~\text{cm}$     \\
                            & Sampling Rate          	& $250~\text{Hz}$     \\ 
                   		& Resolution (Bit Depth) 	& $20~\text{bit}$       \\
                   		& Bandwidth 							& $0.05-50~\text{Hz}$ \\
\cline{1-3}
\multirow{6}{*}{DCG} 	& Carrier Frequency 		& $77-81~\text{GHz}$ \\
										& Sampling Rate          	& $5000~\text{kHz}$     \\ 
                   		& Resolution (Bit Depth) 	& $16~\text{bit}$       \\
                   		& Bandwidth 							& $4~\text{GHz}$ \\
                        & Tx used   & 12 \\
                        & Rx used   & 16 \\
\cline{1-3}
\end{tabular}
\label{tab:system parameter configurations}
\end{table}

\subsection{SCG and ECG Modules}
This subsection introduces the SCG and ECG sensor modules in detail, including the system configurations and hardware design.

The hardware is composed of two primary components: a core board, and a sensor board. 
The two boards are interconnected via a flexible flat cable (FFC), which allows the use of a standardized core board while enabling flexible replacement of different sensor boards for acquiring various types of physiological signals.
The physical dimensions of the customized SCG and ECG sensors are only $3~\text{cm}\times 3~\text{cm}\times 1.5~\text{cm}$.
This compact format ensures convenient placement on the human subject's thoracic surface while supporting full system functionality.
Furthermore, the reduced size helps minimize sensor-induced motion artifacts, thereby enhancing the fidelity and accuracy of the recorded signals.

The core board is build around a low-power microcontroller unit (MCU), the ESP32C6, which integrates 2.4~GHz WiFi capabilities.
Additional components include a CP2102 Universal Serial Bus (USB) to universal asynchronous receiver–transmitter (UART) bridge, a 5~V to 3.3~V linear voltage regulator, and various peripheral interface and support circuits.

The SCG sensor board is primarily composed of an ADXL355 triaxial accelerometer (from Analog Devices). 
This sensor features ultra-low noise density, low 0~g offset drift, and low power consumption.
It includes a 20-bit analog-to-digital converter (ADC) and programmable high-pass and low-pass digital filters.
The ADXL355 supports measurement ranges of $\pm2~\text{g}$, $\pm4~\text{g}$, and $\pm8~\text{g}$, and outputs X, Y, and Z axis acceleration values at up to 4~kHz via either Serial Peripheral Interface (SPI) or Inter-Integrated Circuit ($\text{I}^2\text{C}$) interfaces. 
These characteristics make the sensor well-suited for detecting subtle acceleration changes as required in SCG signal acquisition.
The mechanical vibrations generated by the heart typically range from $20~\text{mG}$ to $40~\text{mG}$, and the minimum time interval between adjacent fiducial points is appoximately $25~\text{ms}$. 
To capture these events with sufficient temporal resolution, the SCG module employs a sampling rate of $250~\text{Hz}$ and a resolution of $20~\text{bit}$, ensuring accurate extraction of signal features.

The ECG sensor board is build around the ADS1292 analog front-end (AFE) chip, which provides two 24-bit ADC channels capable of acquiring standard Lead II ECG signals.
It also integrates feature such as a right-leg drive amplifier, continuous lead-off detection, and internal test signal generation.
The ADS1292 operates over a voltage range of 1.7~V to 3.6~V and is widely used in clinical applications such as patient monitoring, ambulatory ECG, automated external defibrillators (AEDs), and telemedicine.
Normal cardiac mechanical vibrations occur at a frequency of approximately $1~\text{Hz}$ to $2~\text{Hz}$, and the QRS complex typically lasts between $0.06~\text{s}$ and $0.10~\text{s}$, not exceeding $0.11~\text{s}$. 
Accordingly, the ECG module adopts a sampling rate of $125~\text{Hz}$, which is sufficient to accurately capture the QRS complex while minimizing data volume.

Both sensor boards are fabricated using Flexible Printed Circuit (FPC) technology. 
FPCs are manufactured using flexible substrates such as polyethylene terephthalate (PET) or polyimide (PI), offering high reliability and excellent bendability.
This technology embeds electronice circuits into a thin, flexible plastic substrate, enabling dense integration of components within confined spaces.
FPC-based designs offer advantages including lightweight structure, small footprint, superior heat dissipation, and ease of installation.
In this system, the ECG and SCG sensor modules are both implemented on FPC substrates.
The sesors are compact and lightweight, allowing for stable and secure attachment to the chest surface.
Additionally, the flexibility of FPCs enhances conformity to the irregular curvature of human skin, improving contace stability.
Another key advantage of FPCs is their ability to isolate mechanical vibrations originating from the core board, thereby significantly improving the quality of SCG signals during data acquisition.

\subsection{DCG Module}
For the acquisition of DCG signals, our system employs an mmWave radar module based on the AWR2243 chip.
The radar operates within the 76~GHz to 81~GHz frequence band using frequency-modulated continuous-wave (FMCW) signals.
It achieves a range resolution of 3.75~cm. 
The millimeter-scale wavelength enables precise estimation of micromotions along the radial direction by analyzing phase variations in the received signals.
The radar setup consists of a four-chip cascaded transceiver module, which supports up to 12 TX and 16 RX antennas, forming a total of 192 virtual receive channels through MIMO configuration.
This antenna array specifications result in an angular resolution of $1.4\degree$, corresponding to the cross-range resolution of 2.44~cm at a distance of 1.0~m. Such high cross-range resolution facilitates fine-grained spatial detection of the human body, enabling differentiation among multiple auscultation sites. 
Within a suitable sensing range, the radar can simultaneously and reliably capture cardiac mechanical activity at five standard auscultation points, making it well-suited for comprehensive, non-contact cardiac monitoring.

\subsection{Other Considerations}
The system's overall power consumption primarily arises from the active components. 
Based on current consumption under typical operating conditions, the ESP32 core board draws approximately $200~\text{mA}$, the ADXL355 sensor approximately $0.15~\text{mA}$, and the ADS1292 analog front-end approximately $0.2~\text{mA}$. 
This yields a total current draw of about $200.35~\text{mA}$. Operating at a supply voltage of 3.7~V, the system's total power consumption is approximately 0.74~W. 
With a mAh battery, the system achieves a continuous runtime of approximately 3.5~hours.
Due to the hardware limitations of the wireless transceiver module (ESP32-C6), the wireless transmission bandwidth is constrained.
Experimental evaluation using the onboard PCB antenna within a 10-meter range and at full transmission power yielded a peak bandwidth of 25~Mbps, with an average of 23~Mbps. 
Based on computation, each SCG sensor generates a data rate of approximately 5~Kbps, while the ECG sensor generates approximately 2.5~Kbps, resulting in a total transmission load of 27.5~Kbps.
Thus, the system's configurations remain well within the available transmission bandwidth limits. 
\section{Data Processing and Feature Extraction}
\label{sec:data processing methods}

\begin{figure*}[htb]
\centering
\begin{minipage}[t]{0.48\linewidth}
\centering
\includegraphics[width=\linewidth]{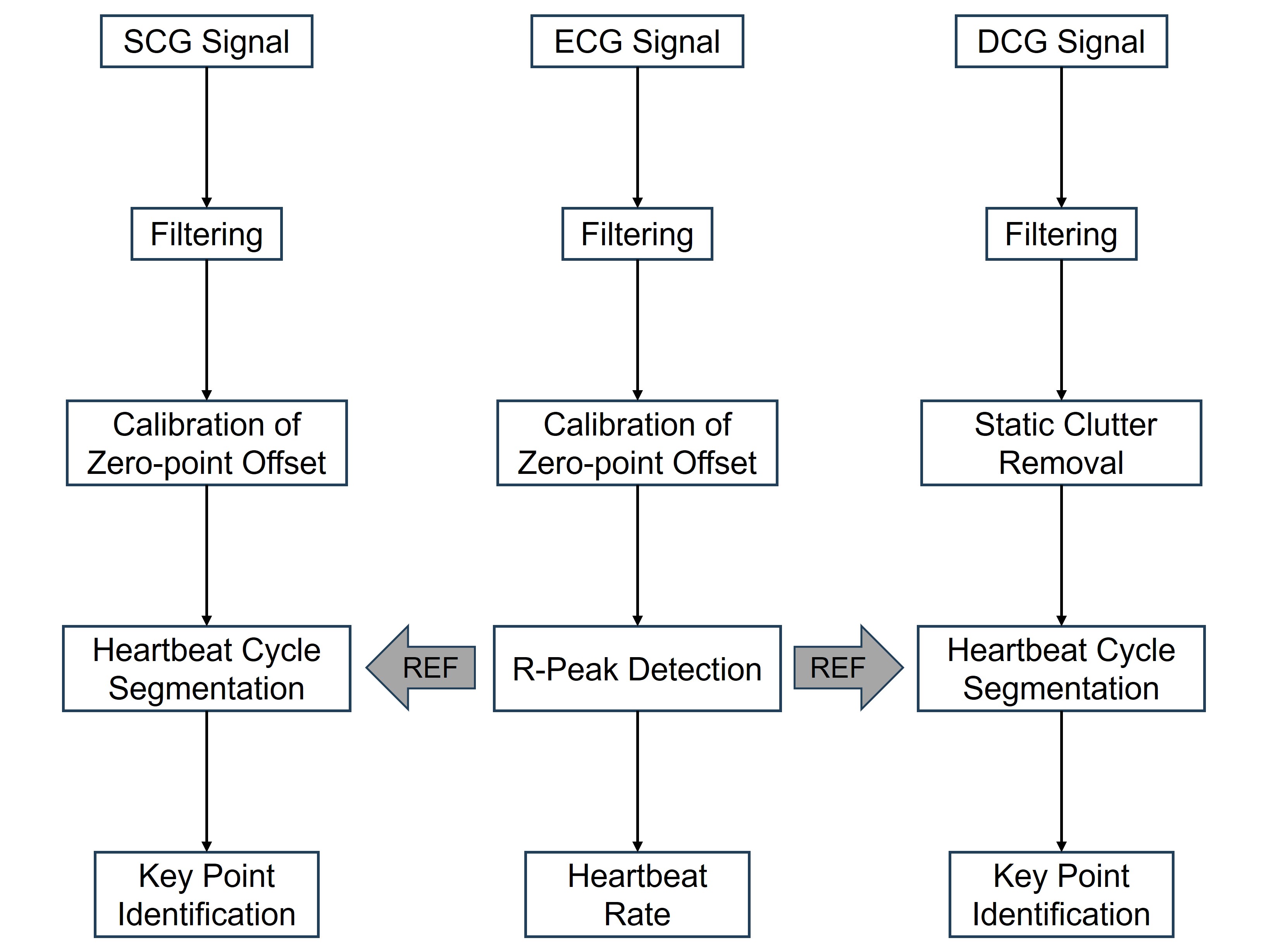}
\caption{Signal-processing pipeline for tri-modality signals. The pipeline comprises preprocessing/denoising/clutter removal, ECG R-peak detection for cycle delineation, cycle segmentation across modalities, and extraction of SCG/DCG key feature points..}
\label{fig:signal processing method}
\end{minipage}\hfill
\begin{minipage}[t]{0.48\linewidth}
\centering
\includegraphics[width=\linewidth]{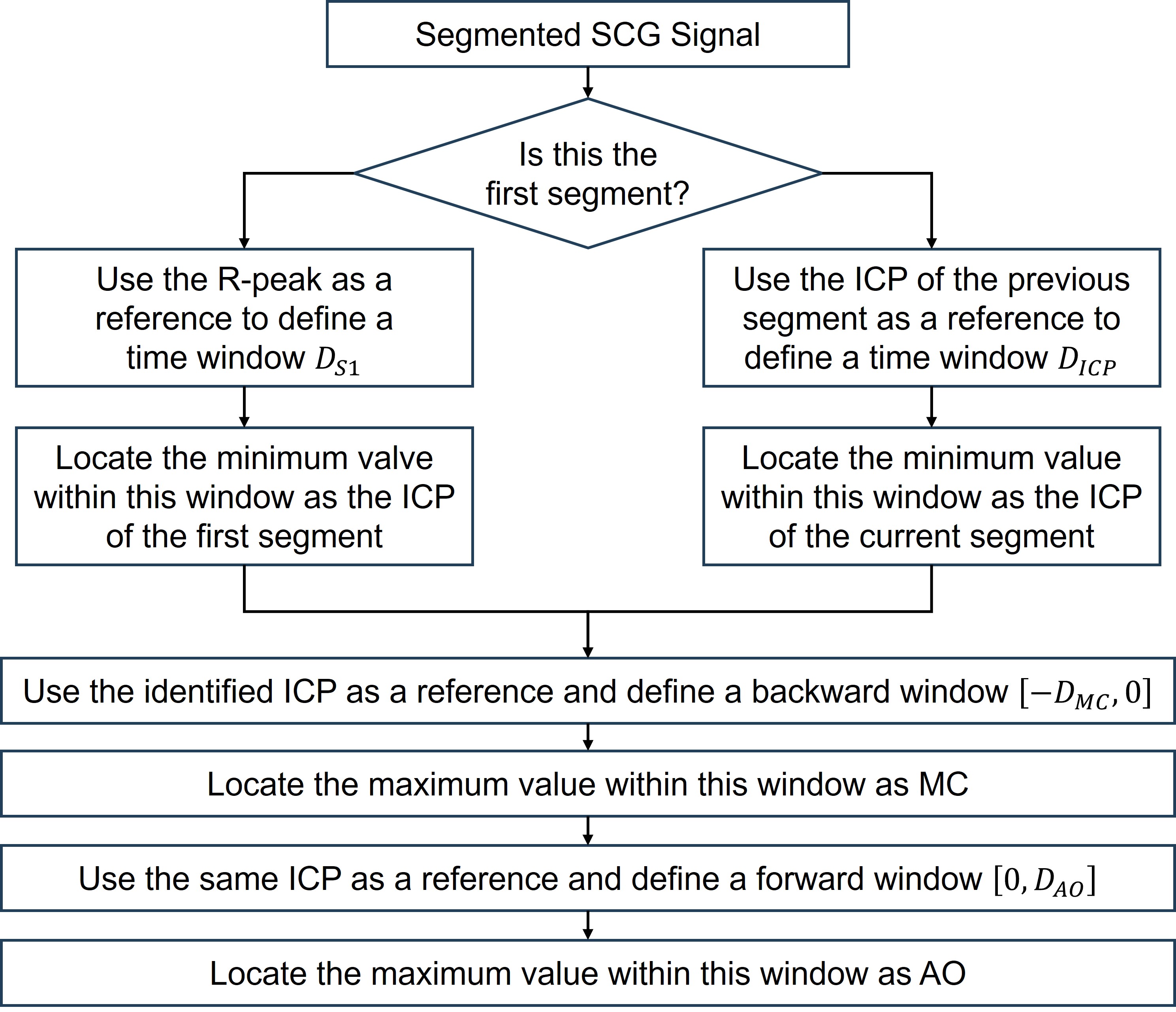}
\caption{SCG key feature point detection procedure. Beat-wise identification of AO, AC, MC, and MO using ECG-anchored windows within S1/S2.}
\label{fig:finding key points}
\end{minipage}
\end{figure*}

This section describes the signal processing workflow, covering the full pipeline form data acquisition to preprocessing, multimodal signal segmentation, and key feature point extraction.
As shown in Fig.~\ref{fig:signal processing method}, the ECG is first processed to provide temporal references, and then the SCG and DCG signals are processed to extract the relevant physiological information. 

\subsection{Preprocessing}
The first step involves digital filtering of the parsed raw data. 
Since the raw signals contain various forms of noise, preprocessing is essential to remove these interferences and produce cleaner signals suitable for subsequent analysis.
Major sources of noise include power-line interference from system operation and biological signals such as eletromyographic activity.
To address this, bandpass Butterworth filters are applied: a $0.8-2$~HZ filter for SCG signals and a $5-40$~Hz filter for ECG signals. 
This effectively eliminates low-frequency drift and high-frequency noise, preserving the key features associated with cardiac events.

After preprocessing, the three signal modalities (ECG, SCG, and DCG) are segmented on a beat-by-beat basis, using the R-peaks in the ECG as temporal anchors.
For each detected heartbeat, corresponding segments from all three modalities are extracted and stacked to enhance signal quality and to facilitate consistent feature extraction.
The segmentation process proceeds as follows:
\begin{itemize}
    \item First, all R-peaks in the ECG signal are determined.
    \item Then, the SCG and DCG signals are temporally aligned to the ECG signal using the acquisition start timestamp.
    \item Finally, based on the time stamps of the R-peaks, all three modalities are segmented into individual cardiac cycles and stored in chronological order.
\end{itemize}

\subsection{Key Feature Point Extraction}
\label{sec:key feature point extraction}
Both SCG and DCG signals are typically composed of two main waveform complexes: S1, generated by cardiac contraction, and S2, resulting from cardiac relaxation.
Within S1 and S2, additional micro vibrations caused by valve movements and blood flow propagation are present.
These vibrations transmit through internal tissues and organs to the thoracic surface, appearing as sequences of acceleration peaks and valleys in the signal.
Each extremum may correspond to a specific physiological cardiac event.

This study focuses on the identification of four key SCG points: AO (Aortic valve opening), AC (Aortic valve closure), MO (Mitral valve opening), and MC (Mitral valve closure). 
These key points provide essential insights into the cardiac mechanical cycle.
To accurately detect these key points, we propose the following algorithm:
\begin{itemize}
    \item For each segmented heartbeat cycle, the signal is divided into S1 and S2 components.
    \item Within the S1 segment, the search starts from the R-peak, designating a time window $D_{S1}$ for locating the minimum point, denoted as Isovolumetric Contraction Point (ICP).
    \item Using ICP as a reference, a backward window $[-D_{MC},0]$ is defined to locate the first local peak, which is identified as MC, and a forward window $[0,D_{AO}]$ is defined to detect the AO peak.
    \item The same procedure is applied to the S2 segment, using a similar approach to extract AC and MO.
\end{itemize}
The complete key feature point identification process for both S1 and S2 is illustrated in Fig.~\ref{fig:finding key points}.
\section{Experiment and Results}
\label{sec:experiment and results}

This section reports experiments evaluating the feasibility and performance of the proposed tri-modality system. 
Since the proposed tri-modality system is to serve as the development and validation system for multi-point vital sign monitoring with radar, it is expected to offer reliable medical reference (i.e., ECG) and mechanical reference (i.e., multiple SCG). 
For this purpose, we designed two experiments. 
First, the consistency between the DCG and SCG modules was assessed by mounting both on a high-precision linear motion platform and check whether their measured displacements are consistent. 
Next, system performance was validated through human-subject experiments on vital sign monitoring
\footnote{Ethics approval and consent to participate: The Ethic Committee of the First Affiliated Hospital of the Fourth Military Medical University approved the study (No. KY20253449-1). The informed consent of all subjects was obtained prior to their participation.}.

\subsection{Benchtop Validation: DCG–SCG Displacement Agreement}

\subsubsection{Experimental Setup}
This part focuses on evaluating the consistency of the developed DCG and SCG modules.

\begin{figure*}
    \centering
    \subfigure[Five SCG sensors]{\includegraphics[scale=0.4]{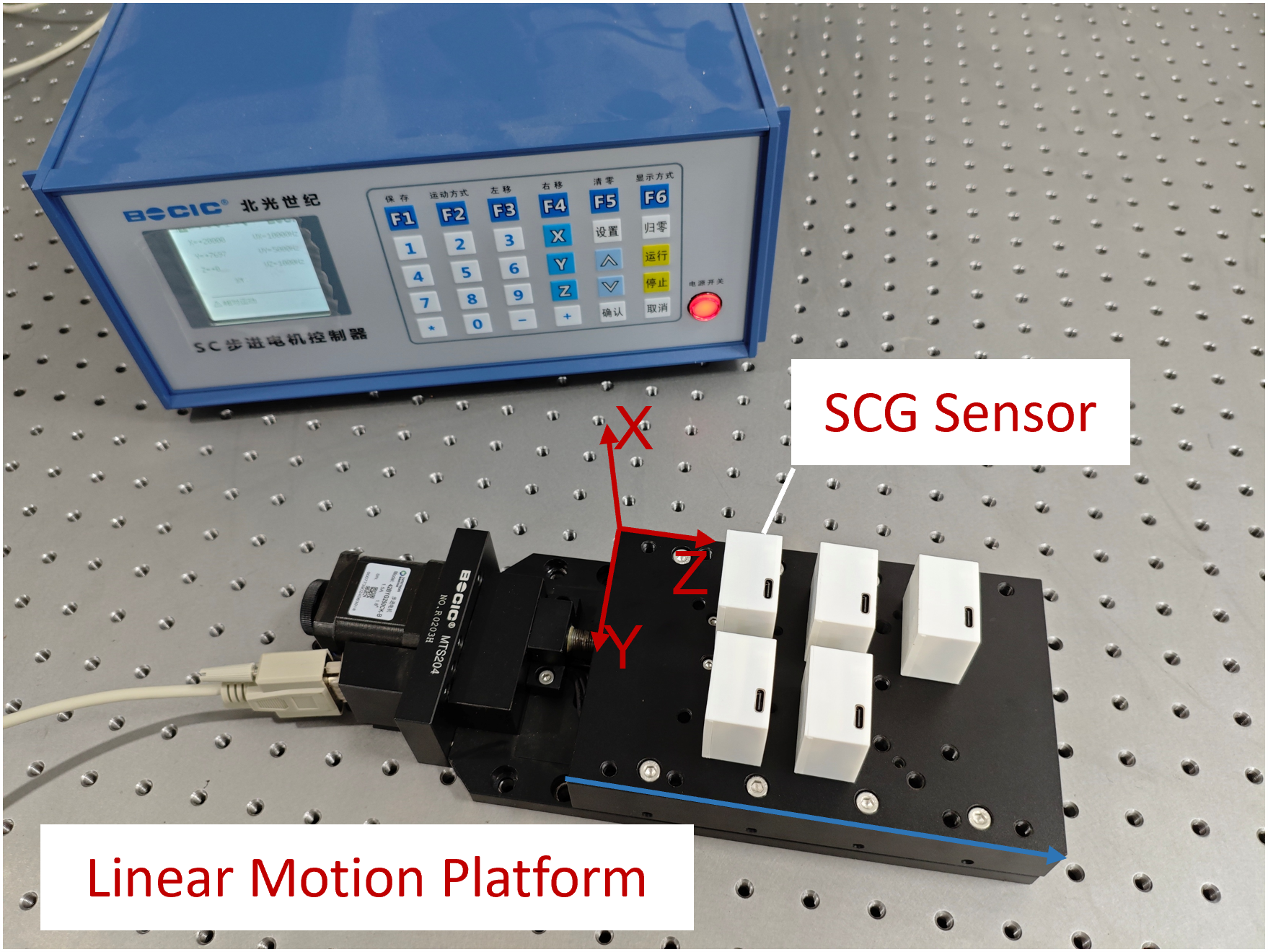}\label{fig:5a}}
    \hfill
    \subfigure[One radar corner]{\includegraphics[scale=0.4]{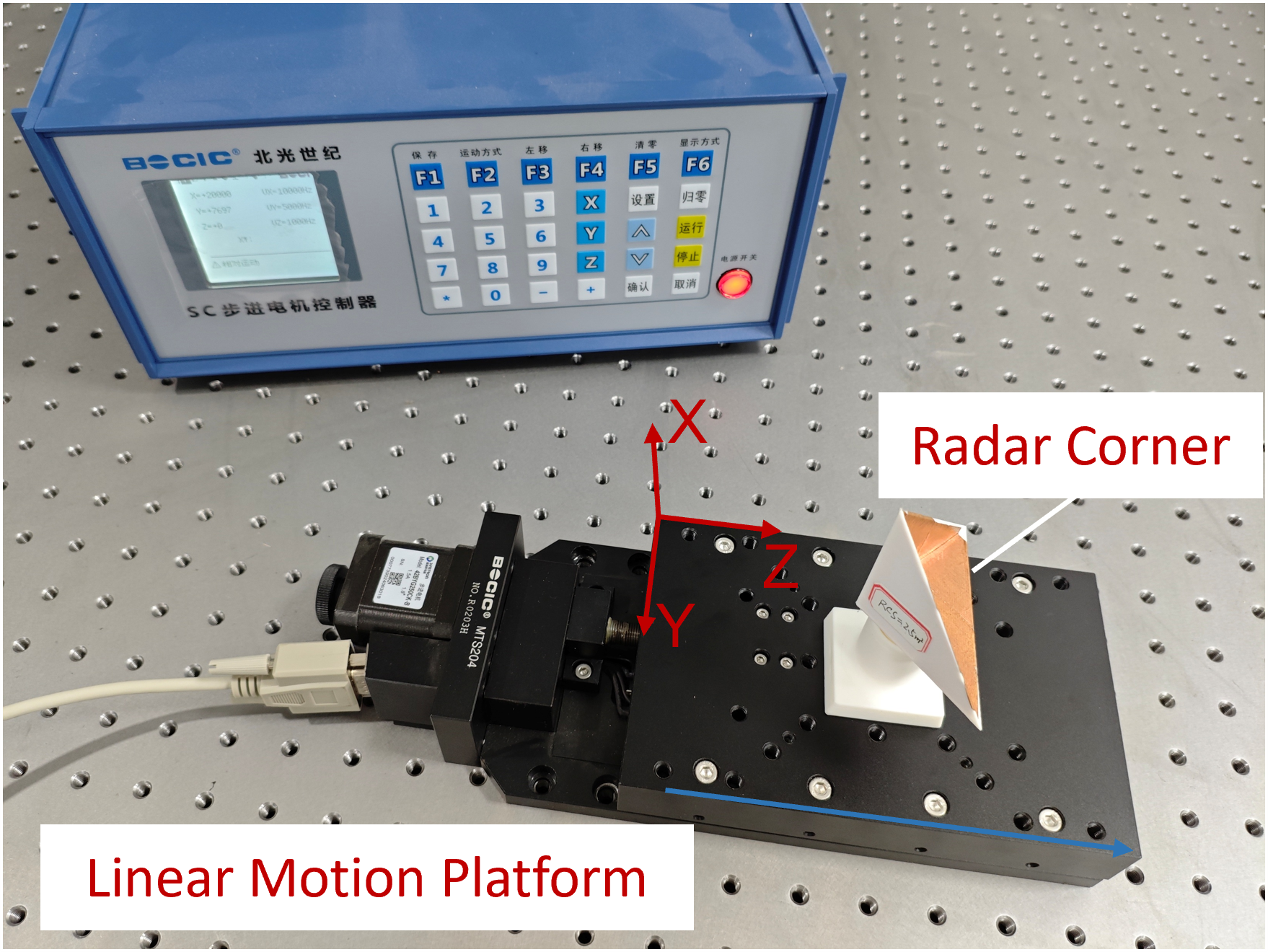}\label{fig:5b}}
    \hfill 
    \subfigure[Five SCG sensors and one radar corner]{\includegraphics[scale=0.4]{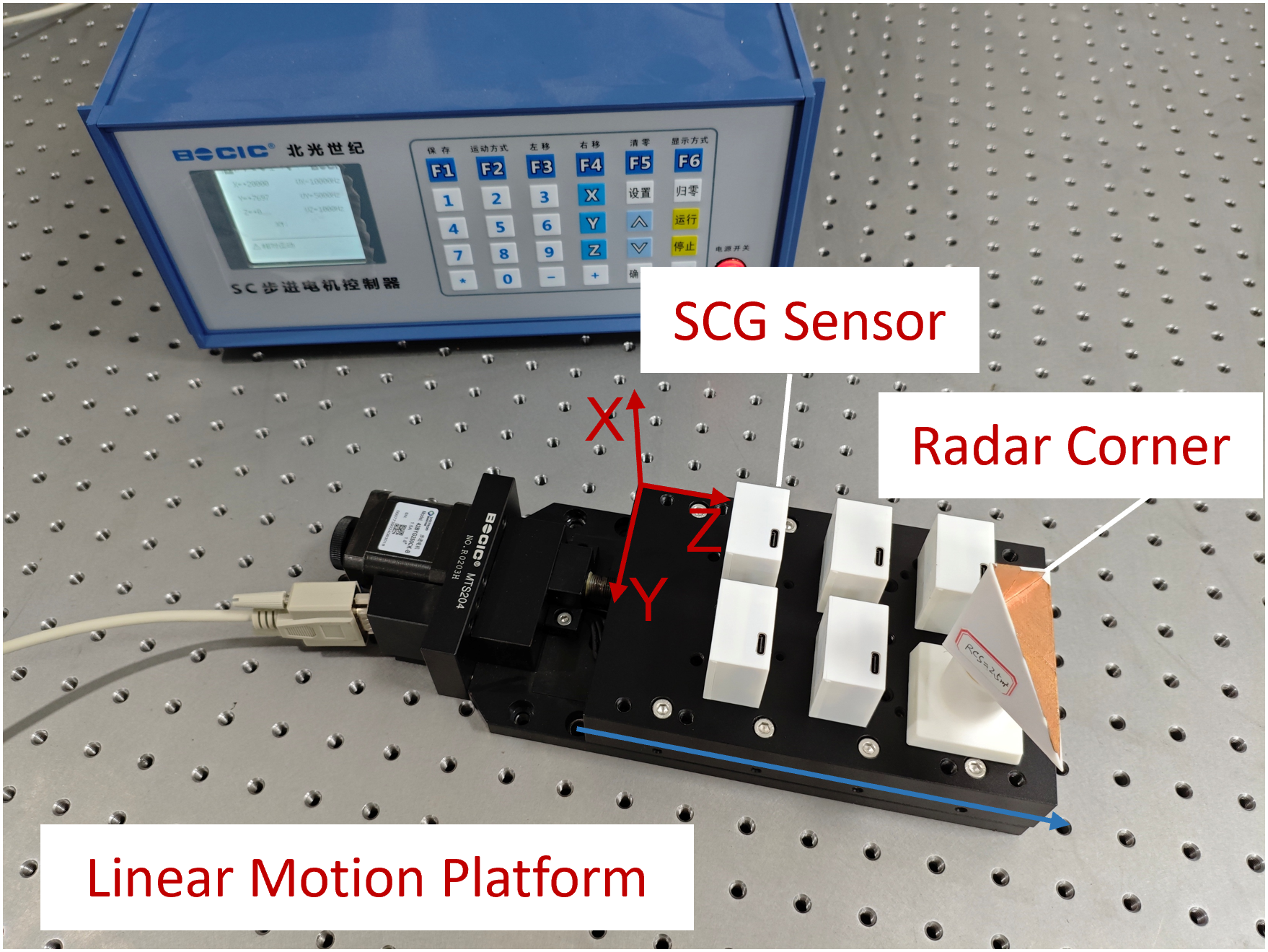}\label{fig:5c}}
    \caption{Benchtop setup for DCG–SCG consistency testing. Five SCG sensors and a radar corner reflector are co-mounted on a programmable linear motion platform with all sensors' Z-axes aligned with the direction of motion. This benchtop provides a controllable mechanical reference to compare DCG-estimated displacement with SCG-derived displacement.}
    \label{fig:sensor modules and linear motion platform}
\end{figure*}

For this purpose, as shown in Fig.~\ref{fig:sensor modules and linear motion platform}.
five SCG sensors were mounted on a computer-controlled linear motion platform, with each sensor's Z-axis aligned with the direction of platform motion.
The platform was programmed to perform linear movements over a fixed distance with varying acceleration parameters.
During the experiment, both the sensor outputs and motion parameters were recorded synchronously.
The detailed procedure is as follows. 
All five SCG sensors were firmly mounted onto the base of the linear motion platform, ensuring mechanical stability and preventing motion artifacts during operation.
The Z-axis of each SCG sensor was carefully aligned with the direction of platform motion.
In addition, a corner reflector having a radar cross section (RCS) of 2.5. is also mounted on the linear motion platform.
The center of the radar's planar antenna array was aligned with the center of the corner reflector, and the normal vector of the antenna plane was oriented parallel to the direction of platform motion.

The motion platform was configured to move linearly from 0~mm to 10~mm, with acceleration increasing from 0~g to 0.2~g during each run.
During each motion cycle, the output data from all five SCG sensors and DCG sensor were collected in real time.
After completing one cycle, the platform returned to its initial position, and the procedure was repeated ten times to ensure repeatability and statistical reliability.

\subsubsection{DCG and SCG Module Performance Evaluation}
The DCG and SCG module performance can be evaluated with the following steps.
\begin{itemize}
    \item Acquire synchronized time-acceleration data from SCG sensors, calculate the displacement of SCG sensors, and calculate the mean value as SCG measured displacement, i.e., $\hat{y}_{SCG}$.
    \item Acquire synchronized radio signals from DCG sensor module, then estimate the displacement of the corner reflector through a series of signal processing steps, i.e., $\hat{y}_{DCG}$.
    \item Compare $\hat{y}_{DCG}$ and $\hat{y}_{SCG}$ qualitatively with a plot and quantitatively with the calculated metric, such as root mean squared error (RMSE).
\end{itemize}

\begin{figure}[htbp!]
\centerline{\includegraphics[scale=0.4]{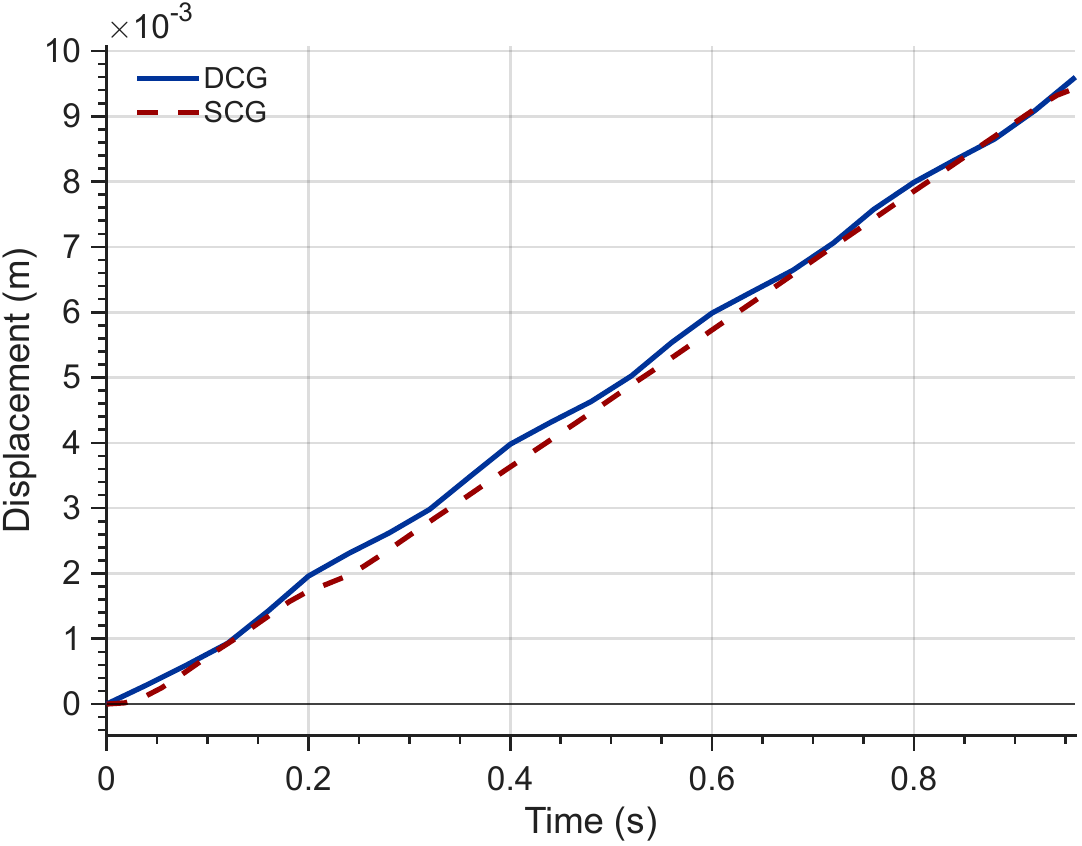}}
\caption{DCG-estimated and SCG-derived displacement on the motion platform. This curves show near-identical trajectories with minor deviations; quantitative metrics indicate RMSE=0.0045~m and Pearson r=0.9947, validating accurate radar displacement estimation relative to the SCG ground truth.}
\label{fig:system performance test DCG vs SCG}
\end{figure}

As shown in Fig.~\ref{fig:system performance test DCG vs SCG}, the displacement curves obtained from the DCG and SCG modules exhibit a high degree of consistency across the entire measurement interval.
Both signals show nearly identical trends with only minor deviations, indicating that the two modalities capture similar mechanical motion.
The close alignment between the DCG and SCG outputs validates the accuracy and reliability of the proposed system in tracking displacement.
Statistical analysis of the data shows that the root mean square error (RMSE) between DCG and SCG displacement is 0.0045~m, and the Pearson correlation coefficient is 0.9947. 
These results confirm that the DCG module provides performance comparable to the established SCG reference. 
This also validates the DCG module's ability to reliably detect subtle mechanical displacements on the human thoracic surface, confirming its suitability for cardiac monitoring applications.

\subsection{Human-Subject Validation: Synchronized Tri-Modality Monitoring}

\begin{table}[]
\caption{Human Subjects' Information}
\centering
\begin{tabular}{lr}
\hline
\textbf{Item} & \textbf{Value} \\ \hline
Sex                        & All male          \\
Age                        & 21,21,20                   \\
Height (cm)                & 178,180,183                \\
Weight (kg)                & 80,77,80                   \\
BMI                        & 25.2, 23.8, 23.9 \\
\hline
\end{tabular}
\label{tab:Subjects' Information}
\end{table}

\begin{figure*}[htb]
\centering
\begin{minipage}[t]{0.45\linewidth}
\centering
\includegraphics[width=\linewidth]{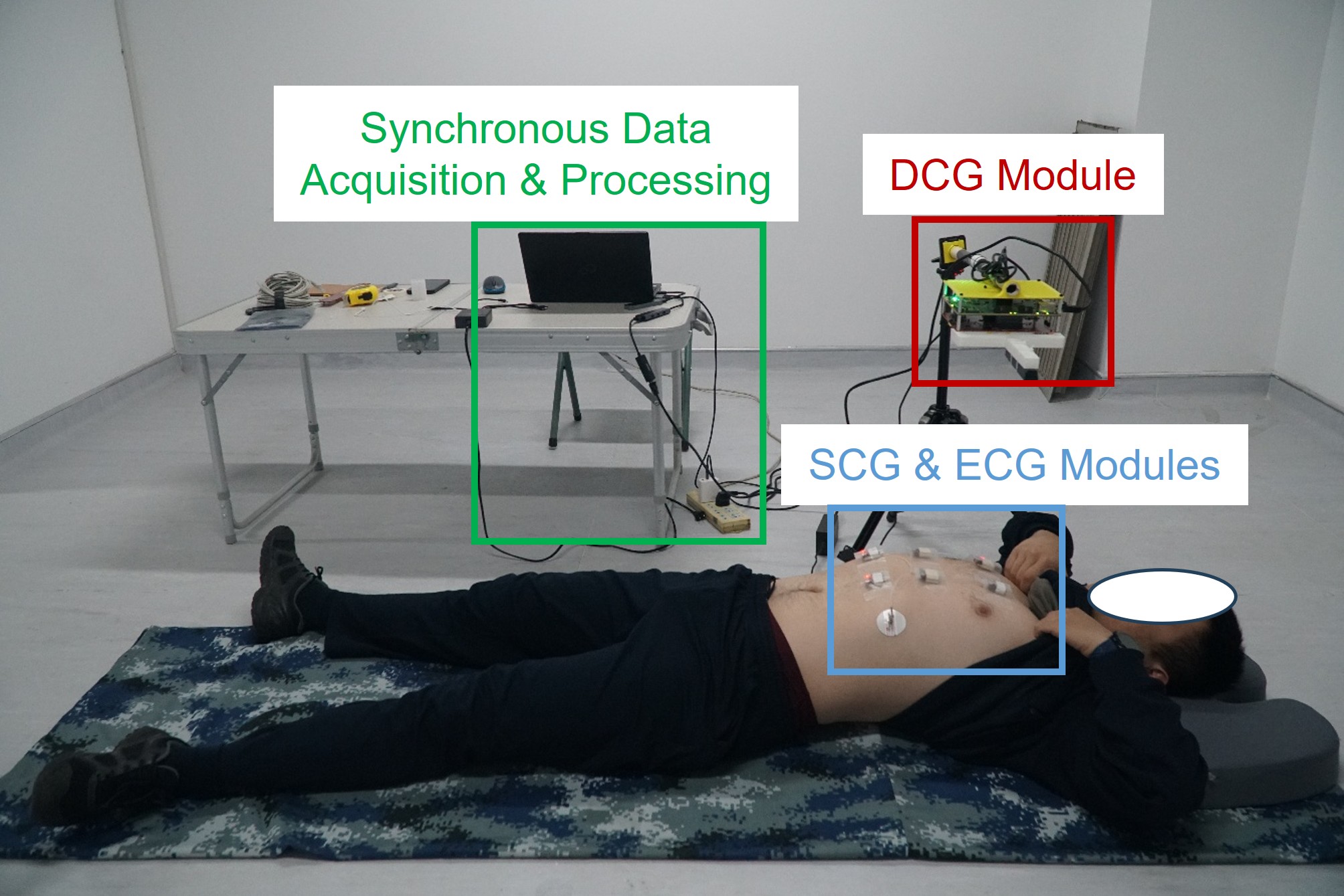}
\caption{Human-subject experimental configuration for synchronized tri-modality recording. An mmWave radar (~0.5 m above the thorax) acquires non-contact DCG, while five SCG sensor modules are affixed at predefined thoracic regions to capture multi-point mechanical signals; Lead-II ECG serves as the electrical anchor. 
}
\label{fig:real experiment scenario}
\end{minipage}\hfill
\begin{minipage}[t]{0.5\linewidth}
\centering
\includegraphics[width=\linewidth]{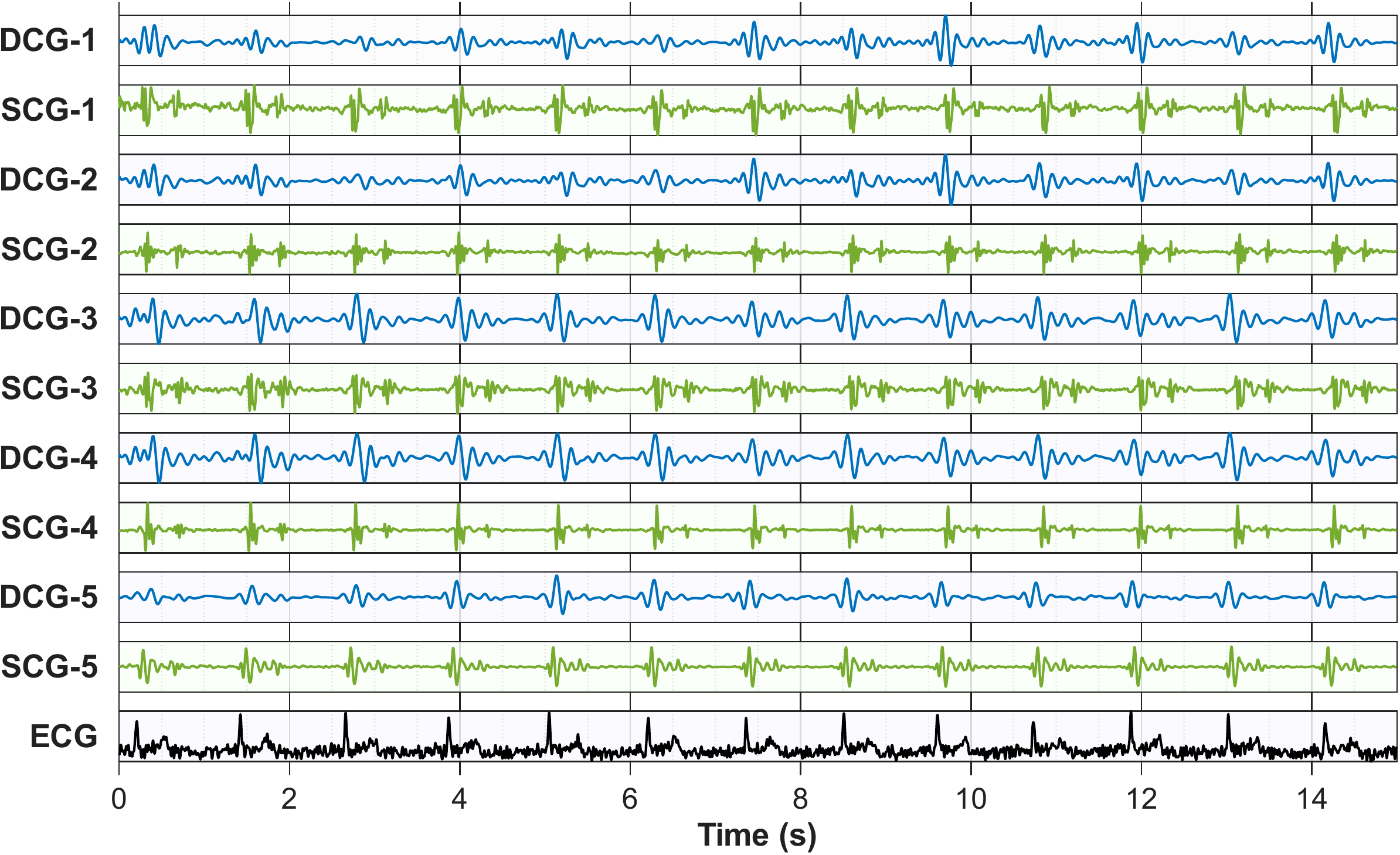}
\caption{Five-channel SCG and DCG with ECG reference (15~s segment). Five-channel SCG and DCG signals were recorded simultaneously along with ECG signals while the subject was in a supine, quasi-static state. }
\label{fig:SCG and DCG signals of 5 sec}
\end{minipage}
\end{figure*}


\subsubsection{Experimental Setup}
This part evaluates the performance of tri-modality sensor system with the experiments on human-subject vital sign monitoring. 

Table.~\ref{tab:Subjects' Information} summarizes the subjects' demographic information in the experiments. All participants were free of cardiovascular disease, with an average age of approximately 21 years.

To investigate the behavior and spatial patterns of SCG signals across multiple channels under different respiratory conditions, a series of controlled experiments were designed.
Subjects were instructed to avoid strenuous physical activity for at least one hour prior to the experiment to minimize heart rate variability and maintain a stable physiological state.
As illustrated in Fig.~\ref{fig:real experiment scenario}, five SCG sensor modules were put on the thorax directly above the five standard cardiac auscultation regions: 
Mitral valve region, Pulmonary valve region, Aortic valve region, Second aortic valve region, and Tricuspid valve region. 
The sensors were secured firmly to the skin to ensure stable contact and eliminate motion artifacts. 
In addition, three disposable ECG electrodes were attached to the left arm, right arm, and left leg to record the ECG signals.
Each subject was required to record approximately 30 seconds of data per trial under supine posture, lying on the floor in a nearly static state without speaking or moving during the data collection period.
Two different respiratory conditions were tested while maintaining this posture:
\begin{itemize}
    \item Natural Breathing: The subject maintained a relaxed, static state and breathed at a self-paced, natural rhythm. Data collection was collected only once the subject was fully prepared. Each recording lasted for 3 minutes.
    \item Breath Holding (Apnea): Prior the start of data collection, the subject took a deep breath and began holding it before recording. The subject maintained breath-holding for 30 seconds, then transitioned to natural breathing for another 30 seconds, and repeated this cycle continuously for a total of 3 minutes. 
\end{itemize}
Each subject was required to complete two experimental trials, with each trial repeated at least twice to ensure consistency and reliability of the data.


\begin{figure*}[htbp]
\centerline{\includegraphics[scale=0.52]{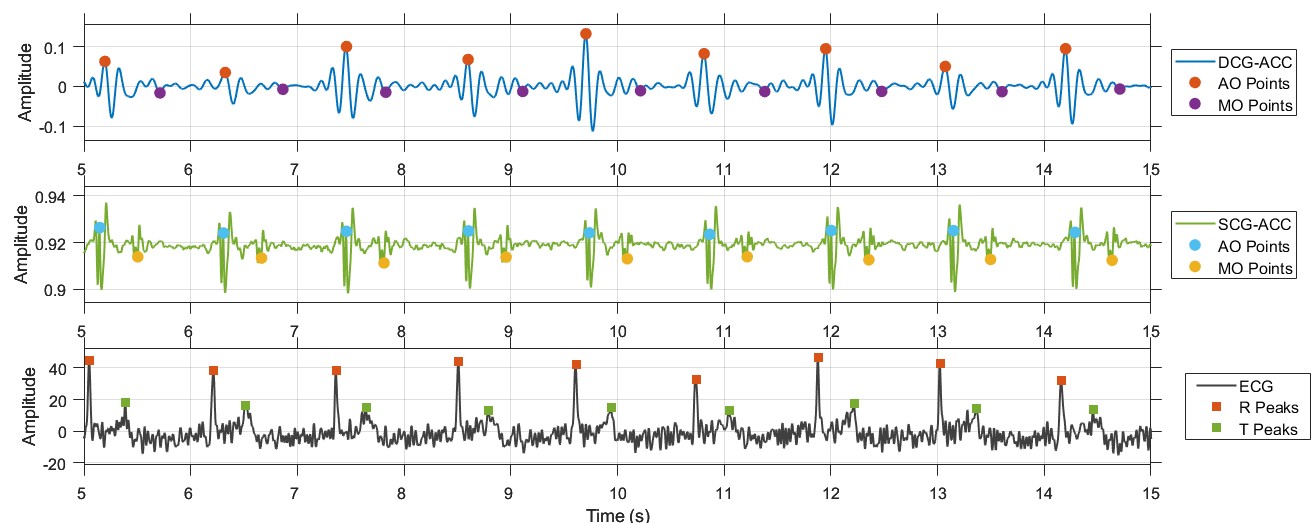}}
\caption{Comparison of SCG, DCG, and ECG signals. ALL key feature points were identified using the proposed feature extraction algorithm and are annotated on the corresponding waveforms.}
\label{fig:SCG_DCG_ECG_signals_and_key_points}
\end{figure*}


\subsubsection{Cross-Modal Key Point Identification}

Based on the SCG key point detection algorithm described in Section~\ref{sec:key feature point extraction}, we conducted a series of experiments to evaluate the applicability and robustness of the method across different signal modalities.
Fig.~\ref{fig:SCG and DCG signals of 5 sec} shows the selected segment of the SCG and DCG signals extracted from the full recording.
Specifically, we selected data from three subjects and extracted several cycles of waveforms of DCG, SCG, and EC signals, as illustrated in Fig.~\ref{fig:SCG_DCG_ECG_signals_and_key_points}.
The key points shown in the figure were automatically labeled using the proposed algorithm and cross-validated against reference SCG signals to ensure correct identification.


The experiment results support the following observations:
\begin{itemize}
  \item \textbf{Coarse scale.} DCG exhibits cycle lengths that are highly consistent with those of SCG and ECG, supporting robust estimation of heart rate and related temporal parameters.
  \item \textbf{Fine scale.} Two complementary perspectives emerge:
  \begin{itemize}
    \item \textit{Intra-modality (across cycles).} The AO--MO interval demonstrates high repeatability within each modality.
    \item \textit{Inter-modality.} Small but systematic discrepancies are observed across modalities in the AO--MO interval (e.g., DCG tends to yield slightly longer intervals than SCG). This bias is plausibly attributable to the displacement-to-acceleration transformation in DCG (DCG-ACC), which amplifies high-frequency noise and can distort the localization of AO and MO fiducial points.
  \end{itemize}
\end{itemize}

\subsection{Discussion}

\begin{table*}[]
\caption{Comparison of the Proposed Tri-modality System with Existing Studies}
\centering
\begin{tabular}{llllll}
\hline
\textbf{Research} & \textbf{Signal Modality} & \textbf{\begin{tabular}[c]{@{}l@{}}Synchronous \\ Monitoring System\end{tabular}} & \textbf{\begin{tabular}[c]{@{}l@{}}Multi-Point \\ Measurement \\ \end{tabular}} & \textbf{\begin{tabular}[c]{@{}l@{}}Key Point \\ Extraction\end{tabular}}
& \textbf{\begin{tabular}[c]{@{}l@{}}Sensor Module \\ Design\end{tabular}} \\ \hline
\cite{shangMultipleObjectLocalization2020}                  & DCG       & No        & No        & No     & No                     \\
\cite{daiEnhancementRemoteVital2022}                        & DCG       & No        & No        & No     & No                     \\
\cite{xuSimultaneousMonitoringMultiple2022}                 & DCG       & No        & No        & No     & No                     \\
\cite{leeResolutionEnhancementTechnique2020}                & DCG       & No        & No        & No     & Yes                     \\
\cite{zhang2024single}                                      & DCG       & No        & Yes       & No     & No                     \\
\cite{helbigSimultaneousElectricalMechanical2016a}          & ECG, DCG  & Yes       & No        & No     & Yes                     \\
\cite{fanRadiofrequencyDopplerEchocardiography2023}         & ECG, DCG  & Yes       & No        & Yes    & No                     \\
\cite{jianRFHeartbeatRobustContactless}                     & ECG, DCG  & No        & No        & No     & No                     \\
\cite{sakamotoNoncontactMeasurementHuman2019}               & ECG, DCG  & No        & No        & No     & No                     \\
\cite{liMotionRobustContactlessHeartbeat2023a}              & ECG, DCG  & No        & Yes       & No     & No                     \\
\cite{munckVisualizationMultichannelSeismocardiogram2019}   & ECG, DCG  & No        & Yes       & No     & No                     \\
\cite{huiMultiPointNearFieldRF2020a}                        & ECG, DCG  & Yes       & Yes       & No     & Yes                     \\
\cite{chenContactlessElectrocardiogramMonitoring2024d}      & ECG, DCG  & Yes       & Yes         & Yes    & No                    \\ 
\cite{gaoMeasurementComplexHuman2019a}                      & ECG, SCG  & Yes       & Yes       & No     & No                     \\
\cite{linIdentificationLocationSpecific2018}                & ECG, SCG  & Yes       & Yes       & Yes    & Yes                     \\
Proposed   &ECG, SCG, DCG                                   & Yes                                                                                  & Yes                                                                                 & Yes                              & Yes                       \\ \hline
\end{tabular}
\label{table:comparison with existing studies}
\end{table*}

\subsubsection{Comparison With Existing Studies}
This section compares the proposed work with existing studies across several key dimensions: signal modality, synchronous monitoring capabilities, multi-point measurement, key point extraction, and sensor design. 
After that a discussion of the current limitations of our system is presented.

In this study, we present a tri-modality wireless system that integrates ECG, SCG, and DCG to address the lack of reliable validation systems for comprehensive cardiac physiological monitoring. 
The system supports synchronous acquisition of all three modalities, enables multi-point fine-grained measurement, and provides detailed information on both sensor design and the key point extraction algorithm.

In previous studies, researchers have explored cardiac status monitoring by integrating either SCG or DCG with ECG. 
A lot of work has focused on the development of DCG systems and signal processing algorithms to enhance the accuracy of non-contact cardiac monitoring \cite{shangMultipleObjectLocalization2020, daiEnhancementRemoteVital2022, xuSimultaneousMonitoringMultiple2022, leeResolutionEnhancementTechnique2020}. 
However, in many of these studies, simple reference systems such as respiration belts or finger-mounted pulse oximeters were employed to provide ground truth data. 
These references are limited in scope and typically only support coarse-level measurements, such as heart rate estimation, without enabling fine-grained physiological analysis.

To address this limitation, several studies have incorporated ECG systems as reference signals to support more precise physiological measurements \cite{chenContactlessElectrocardiogramMonitoring2024d, helbigSimultaneousElectricalMechanical2016a, fanRadiofrequencyDopplerEchocardiography2023, jianRFHeartbeatRobustContactless, sakamotoNoncontactMeasurementHuman2019, liMotionRobustContactlessHeartbeat2023a, munckVisualizationMultichannelSeismocardiogram2019, huiMultiPointNearFieldRF2020a}. 
In most existing ECG-DCG integrated systems, the DCG signal is processed in a two-step pipeline: first estimating body surface motion signal, and then extracting physiological parameters from that motion signal. 
However, these systems lack direct ground truth data for validating DCG outputs, which limits their effectiveness in evaluating and improving radar-based physiological monitoring methods.

In contrast, our study draws from research on ECG-SCG systems \cite{linIdentificationLocationSpecific2018, gaoMeasurementComplexHuman2019a} and advances the field by developing a tri-modality system. 
This approach provides both detailed physiological signals and a direct mechanical reference (SCG) for validating and improving DCG-based techniques. 
The proposed system enables more rigorous validation of radar sensing algorithms and supports more detailed investigation of cardiac physiological monitoring.

A comprehensive comparison between the proposed tri-modality system and existing studies is presented in Table~\ref{table:comparison with existing studies}.

\subsubsection{Limitations}
Although the proposed tri-modality system demonstrates considerable advancements, it still faces two major limitations: the requirement for stationary subjects during data acquisition, and the need for manual spatial calibration between the DCG system (radar antenna arrays) and the SCG sensor positions.

In the current experimental setup, subjects were asked to remain stationary and avoid any body movement. 
This constraint arises from the sensitivity of DCG-estimated signal quality to motion artifacts, which can significantly degrade the accuracy of micro-motion estimation.
To enhance the robustness of this approach in practical scenarios, it is necessary to introduce advanced motion artifact cancelation algorithms.

Another limitation lies in the need for manual spatial calibration, that is, the manual alignment of the radar antenna array relative to the SCG sensors.
This calibration is important to ensure accurate spatial correspondence between the mechanical reference points captured by SCG and the electromagnetic reflections measured by DCG. 
However, in real-world applications, it is impractical to maintain a fixed spatial configuration between the radar system and the subject's torso.
Therefore, to facilitate real-world deployment, future research should incorporate the techniques for the automatic identification of relevant auscultation regions.


\section{Conclusion}
\label{sec:conclusion}

In this work, we present the first tri-modality wireless system that integrates ECG, SCG, and DCG to enable synchronous and fine-grained monitoring for cardiac physiological activity.
This work addresses a critical gap in the field: the absence of a reliable validation platform for evaluating non-contact radar-based sensing techniques.


We presented a research-grade tri-modality wireless platform that integrates ECG, SCG, and DCG under a unified acquisition-and-control framework for time-aligned, multi-point physiological monitoring. 
The system addresses a critical gap by providing both an electrical anchor (ECG) and a direct mechanical ground truth (SCG) against which radar-derived displacment can be rigorously validated, while enabling spatially resolved sensing and cross-modal feature extraction.
Through benchtop validations on a programmable linear-motion platform and human-subject experiments, we demonstrated accurate displacement tracking, robust multimodal signal acquisition, and consistent identification of cardiac feature points (e.g., AO/MO) across modalities. 
Relative to prior work, the platform advances the state of the art by combining synchronous tri-modality acquisition, multi-point thoracic measurement, and a feature point extraction pipeline within a single and wearable system.

Current limitations include the need for stationary subjects and manual spatial calibration between the radar array and SCG sensor locations. 
These limitations primarily reflect DCG’s sensitivity to motion artifacts and the difficulty of maintaining fixed geometry in practical settings. 
Future work will focus on motion-artifact suppression, automatic region-of-interest localization on the thorax, and system calibration to facilitate real-world deployment.

Overall, this tri-modality system establishes a rigorous, spatially resolved validation platform for non-contact physiological sensing and provides a foundation for next-generation radar-based monitoring and algorithm development.

\bibliographystyle{IEEEtran}
\bibliography{main}

\begin{IEEEbiography}
[{\includegraphics[width=1in,height=1.25in,clip,keepaspectratio]{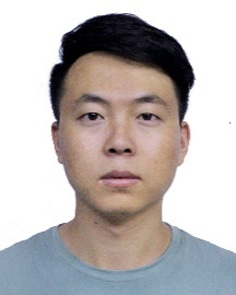}}]
{Wei Ren}
(Member, IEEE) received the B.S. degree from the Department of Precision Instrument, Tsinghua University, Beijing, China, in 2015, and Ph.D. degree from School of Electronics and Information, Beijing Institute of Technology, China, in 2021. He was also a visiting student with the Department of Electrical Engineering and Computer Science, The University of Tennessee, Knoxville, TN, USA from 2019 to 2020. 

He is currently working as a Postdoc Fellow with 
the Hong Kong University of Science and Technology (HKUST).
His research interests include wireless human sensing, radar waveform design, radar signal processing, and machine learning.
\end{IEEEbiography}

\begin{IEEEbiography}
[{\includegraphics[width=1in,height=1.25in,clip,keepaspectratio]{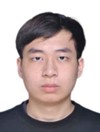}}]
{Huansheng Yi} received the B.E. degrees in internet of things engineering from Luoyang Institute of Science and Technology in 2023, Henan, China.

He is currently serving as a RA in the BME department, FMMU, Xi’an. Since 2019, He majored in internet of things and embedded technology, school of computer and information engineering, Luoyang, Henan, China. He won the third prize in the ICAN Innovation Contest(final), 2021. He won the second prize of China Undergraduate Computer Design Competition, 2022. He won the third prize of China Undergraduate Service Outsourcing Innovation and Entrepreneurship Competition, 2022. He also awarded scholarship for many times in college. 
\end{IEEEbiography}



\begin{IEEEbiography}
[{\includegraphics[width=1in,height=1.25in,clip,keepaspectratio]{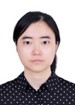}}]
{Miaoyang Hu} received a MBBS and Ph.D. degrees in clinical medicine (eight-year program) in 2014 and 2016 from Wuhan University , Wuhan, China.

Since 2016, she has been an attending physician at department of cardiology, Xijing Hospital, Fourth Military 
Medical University (FMMU) and engaged in clinical cardiac electrophysiology. She has rich clinical practical experience in the diagnosis and interventional treatment of various types of tachyarrhythmias and bradycardiac arrhythmias. In July 2023, she went to study CRT implant and programming optimization in Westmead Hospital, Sydney, NSW. In October 2023, she  went to study cardiac electrophysiology in Nanjing Medical University, Nanjing, China. She has published over 10 peer-reviewed journal and conference papers. Her research interest is cardiac electrophysiology.
\end{IEEEbiography}

\begin{IEEEbiography}
[{\includegraphics[width=1in,height=1.25in,clip,keepaspectratio]{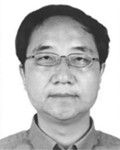}}]
{Jianqi Wang} was born in Xi’an, Shaanxi, China, in 1962. He received the B.E. degree in information and control engineering from Xi’an Jiaotong University, Xi’an, in 1984; the M.E. degree in electronic engineering from the National University of Defense Technology, Changsha, China, in 1990; and the Ph.D. degree from the Key Laboratory of the Ministry of Education of China, Xi’an Jiaotong University, in 2006.

Since 1990, he has taught in the School of Biomedical Engineering, Fourth Military Medical University, Xi’an. He is currently serving as a Professor with and the Director of the Department of Electronics. He pioneered radar-based human being detection in China in 1998 and has published more than 100 articles on the technology. His research interest is bio-radar technology, including signal processing, human being detection, and imaging.
\end{IEEEbiography}

\begin{IEEEbiography}
[{\includegraphics[width=1in,height=1.25in,clip,keepaspectratio]{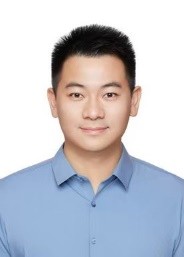}}]
{Fugui Qi} (Member, IEEE) received the B.E. and Ph.D. degrees in biomedical engineering from the Institute of Biomedical Engineering (BME), Fourth Military Medical University (FMMU), Xi’an, China, in 2014 and 2020, respectively.

Since 2020, he has been with the School of Biomedical Engineering, Fourth Military Medical University, Xi’an.  He is currently serving as a research fellow in the Bio-radar and Signal Processing Laboratory, School of BME, FMMU.  From 2019 to 2020, he was a visiting scholar with The University of Tennessee, Knoxville, TN, USA.  He has published more than 40 peer-reviewed journal and conference papers.  His research interests include bio-radar-based medical signal processing, through-wall human motion recognition, and noncontact intelligent life sensing. He received the Excellent Doctoral Dissertations of Shaanxi Province in 2022 and was selected for the Young Talent Fund of the Association for Science and Technology in Shaanxi, China, in 2023. He also won the First Prize for Young Excellent Paper Competition (Young Scholar) of the 2023 China BME conference and the Second Award for Science Research Famous Achievement Award in Shanxi Higher Institution in 2024.

\end{IEEEbiography}

\end{document}